\documentclass[final,5p,times,twocolumn]{elsarticle}
\biboptions{numbers,sort&compress}

\usepackage{amssymb}
\usepackage{amsmath}
\usepackage{graphicx}
\usepackage{color}
\usepackage{soul}

\journal{Ultramicroscopy. Available online, 4 Feb 2019, doi:10.1016/j.ultramic.2019.02.003}
\tnotetext[a1]{\textsuperscript{\textcopyright } 2019. This manuscript version is made available under the CC-BY-NC-ND 4.0 license http://creativecommons.org/licenses/by-nc-nd/4.0/}

\begin{document}

\begin{frontmatter}

\title{The maximum a posteriori probability rule for atom column detection from HAADF STEM images}

\author[label1,label2]{J. Fatermans}
\author[label1]{S. Van Aert\corref{cor}}
\ead{Sandra.VanAert@uantwerpen.be}
\cortext[cor]{Corresponding author}
\author[label2,label3]{A.J. den Dekker}

\address[label1]{Electron Microsopy for Materials Science (EMAT), University of Antwerp, Groenenborgerlaan 171, 2020 Antwerp, Belgium}
\address[label2]{imec-Vision Lab, University of Antwerp, Universiteitsplein 1, 2610 Wilrijk, Belgium}
\address[label3]{Delft Center for Systems and Control (DCSC), Delft University of Technology, Mekelweg 2, 2628 CD Delft, Netherlands}

\begin{abstract}
Recently, the maximum a posteriori (MAP) probability rule has been proposed as an objective and quantitative method to detect atom columns and even single atoms from high-resolution high-angle annular dark-field (HAADF) scanning transmission electron microscopy (STEM) images. The method combines statistical parameter estimation and model-order selection using a Bayesian framework and has been shown to be especially useful for the analysis of the structure of beam-sensitive nanomaterials. In order to avoid beam damage, images of such materials are usually acquired using a limited incoming electron dose resulting in a low contrast-to-noise ratio (CNR) which makes visual inspection unreliable. This creates a need for an objective and quantitative approach. The present paper describes the methodology of the MAP probability rule, gives its step-by-step derivation and discusses its algorithmic implementation for atom column detection. In addition, simulation results are presented showing that the performance of the MAP probability rule to detect the correct number of atomic columns from HAADF STEM images is superior to that of other model-order selection criteria, including the Akaike Information Criterion (AIC) and the Bayesian Information Criterion (BIC). Moreover, the MAP probability rule is used as a tool to evaluate the relation between STEM image quality measures and atom detectability resulting in the introduction of the so-called integrated CNR (ICNR) as a new image quality measure that better correlates with atom detectability than conventional measures such as signal-to-noise ratio (SNR) and CNR. 
\end{abstract}

\begin{keyword}
Scanning transmission electron microscopy (STEM) \sep Atom detection \sep Atom detectability \sep Model selection
\end{keyword}

\end{frontmatter}

\section{Introduction}\label{Introduction}
The physical properties of nanomaterials are strongly related to their exact structural and chemical composition. Small changes in the local atomic structure may induce significant changes in their properties \cite{qi(2010),alem(2011),vanaert(2012)2,tang(2014)}. Therefore, precisely measuring the atomic arrangement of projected atomic columns or individual atoms is important in order to fully understand the structure-properties relation of nanomaterials. Scanning transmission electron microscopy (STEM) has become a widely used technique to visualize nanomaterials with sub-angstrom resolution due to improvements in aberration correction technology \cite{rose(2009),hawkes(2015)}. The high-angle annular dark-field (HAADF) regime is considered predominantly incoherent \cite{loane(1992),hartel(1996),nellist(1999)}, allowing for directly interpretable images due to the lack of contrast reversals. However, a merely visual interpretation of HAADF STEM images is inadequate to obtain precise structure information. A quantitative approach is required which can be provided by statistical parameter estimation \cite{dendekker(2005),vanaert(2005),vandenbos(2007),vanaert(2012)1,dendekker(2013),debacker(2016)}, enabling one to determine the atomic column positions \cite{vanaert(2012)2,kundu(2014),akamine(2015),gonnissen(2016)}, chemical composition \cite{vanaert(2009),martinez(2014)} or the number of atoms in an atomic column \cite{bals(2012),vanaert(2013)}. When atom counting results, obtained from two-dimensional (2D) STEM images, are combined from different viewing directions, a three-dimensional (3D) reconstruction of the material at the atomic level is achievable \cite{vanaert(2011),bals(2011),vandenbos(2016)}. Even from a single projection image, 3D atomic models can be obtained \cite{debacker(2017)}. \\
\indent In statistical parameter estimation theory, STEM images are considered as data planes from which unknown structure parameters are estimated. For this, a parametric model is needed describing the expectations of the experimental measurements. Quantitative structure information is obtained by fitting the model to the experimental images using a criterion of goodness of fit. For atomic resolution STEM images, the projected atomic columns can be described as Gaussian functions superimposed on a constant background \cite{vandyck(2002),nellist(2007),debacker(2016)}. An important assumption in this quantitative approach is that the number of atomic columns is known. Usually, this number is determined visually, which is possible for atomic resolution images of beam-stable materials where a high incoming electron dose can be used resulting in images exhibiting high signal-to-noise ratio (SNR). However, beam-sensitive materials should be imaged with a sufficiently low electron dose to avoid beam damage \cite{williams(1996)}. As a consequence, these images exhibit low SNR and low contrast and hence low contrast-to-noise ratio (CNR). This causes poor visual determination of the number of atomic columns in the image leading to biased structure information. To overcome this problem, an alternative, quantitative method has been proposed in a recent study to determine the number of atomic columns for which there is most evidence in the image data \cite{fatermans(2018)}. The method, which is referred to as the maximum a posteriori (MAP) probability rule, is based on a combination of statistical parameter estimation and model-order selection. It has been shown to be able to automatically and objectively determine the most probable structure of unknown nanomaterials and to detect single atoms with high reliability. Moreover, it quantifies how more likely an obtained atomic structure is as compared to other structures. The present paper covers a detailed derivation of an approximate analytical expression for the MAP probability using a Bayesian approach. In addition, it is shown that the proposed MAP probability rule is a model-order selection criterion and its relation to other criteria is discussed together with a comparison of the performance in detecting atoms from HAADF STEM images. Moreover, since atom detectability is related to image quality, the MAP probability rule can be used as a tool to investigate the relation between image quality measures and atom detectability. As such, in this work, a new image quality measure is proposed that better correlates with atom detectability than conventional image quality measures. \\ 
\indent The remainder of this article will be organised as follows. In section \ref{Methodology}, the methodology of the MAP probability rule to determine the most probable number of atomic columns from HAADF STEM images is described in detail. This is followed in section \ref{quality} by an explanation of  how the MAP probability rule can be used to evaluate the connection between measures of image quality and atom detectability. In section \ref{modelselection}, the relation of the MAP probability rule to model selection is investigated. Finally, in section \ref{conclusions}, conclusions are drawn.

\section{Methodology}\label{Methodology}

\subsection{Model-based parameter estimation}\label{Model-based parameter estimation}
Quantitative measurements of the unknown structure parameters, such as atomic column positions and their widths, peak intensities and background, are not directly provided by high-resolution STEM images, but can be obtained from these images using statistical parameter estimation theory \cite{dendekker(2005),vanaert(2005),vandenbos(2007),vanaert(2012)1,dendekker(2013),debacker(2016)}. The starting point of this procedure is the construction of a parametric physics-based model that describes the expectations of the image pixel values as a function of the unknown structure parameters. This model is then fitted to the observed image pixel values with respect to the unknown structure parameters using a criterion of goodness of fit. Ideally, the model accurately describes the image formation process required for a STEM computer simulation, including dynamical electron diffraction effects, thermal diffuse scattering, electron-sample interaction, microscope transfer function and detector efficiency. However, since model parameters are estimated by an iterative optimisation scheme, using this type of complex models becomes very time consuming. For this reason, often a simplified empirical model is used. For STEM images, the intensity is sharply peaked at the atomic column positions \cite{vandyck(2002),nellist(2007)}. Therefore, the observed STEM pixel values of images of K$\times$L pixels, $\textbf{w}$ = (w$_{11}$, ..., w$_{\text{KL}}$)$^\text{T}$, can be modelled as a superposition of Gaussian peaks \cite{debacker(2016)}. When a different width is assumed for each estimated Gaussian peak, the expectation model f$_{kl}$($\boldsymbol{\theta}$), with $\boldsymbol{\theta}$ the vector of unknown structure parameters, describes the expectation of the observed pixel value w$_{kl}$ at position (x$_k$,y$_l$):
\begin{equation}
\label{eq:model}
f_{kl}(\boldsymbol{\theta}) =\zeta+\sum_{n=1}^{N}\eta_n exp\bigg(-\frac{(x_k-\beta_{x_n})^2+(y_l-\beta_{y_n})^2}{2\rho_n^2}\bigg)
\end{equation}
where $\zeta$ is a constant background, $\rho_n$, $\eta_n$, $\beta_{x_n}$ and $\beta_{y_n}$ are the width, the height and x- and y-coordinates of the nth atomic column described by a Gaussian peak, respectively, and N is the total number of atomic columns. The unknown parameters of the expectation model are represented by the parameter vector:
\begin{equation}
\label{eq:parametervector}
\boldsymbol{\theta}=(\beta_{x_1},...,\beta_{x_N},\beta_{y_1},...,\beta_{y_N},\rho_1,...,\rho_N,\eta_1,...,\eta_N,\zeta)^T
\end{equation}
containing M = 4N+1 parameters.

\subsection{Bayesian approach}
In order to extract reliable structure information of nanomaterials from HAADF STEM images using a model such as in Eq. (\ref{eq:model}), the number of atomic columns N present in the image should be known. Usually, for beam-stable materials this number can be determined visually due to the high incoming electron dose that can be used to image these materials, leading to images with high SNR. For beam-sensitive nanostructures, though, the incoming electron dose is limited in order to avoid beam damage and the images exhibit low SNR and low contrast and hence low CNR. Visual inspection of such images may lead to biased results. To overcome this problem, the number of atomic columns N can be reliably quantified by the recently proposed MAP probability rule, which is a combination of statistical parameter estimation and model-order selection \cite{fatermans(2018)}. It selects the number of columns N of which the posterior probability given the observed image pixel values \textbf{w}, p(N$\vert$\textbf{w}), is maximised. By using Bayes' theorem \cite{sivia(2006)}, p(N$\vert\textbf{w}$) can be written as 
\begin{equation}
\label{eq:MAP}
p(N\vert\textbf{w}) = \frac{p(\textbf{w}\vert N)p(N)}{p(\textbf{w})}.
\end{equation}
The term p(\textbf{w}$\vert$N) reflects the evidence that the image data \textbf{w} is generated by N atomic columns. The probability p(N) expresses prior knowledge of the number of atomic columns N in the image, which has been chosen to be a uniform distribution, reflecting no a priori preference for any number of columns. The denominator is a normalization constant, which is independent of the number of columns N, and therefore cancels out when comparing posterior probabilities as a function of N. As such, Eq. (\ref{eq:MAP}) reduces to
\begin{equation}
\label{eq:MAPred}
p(N\vert\textbf{w}) \propto p(\textbf{w}\vert N).
\end{equation}
By making use of so-called marginalisation \cite{jaynes(2003)}, which can be considered as integrating out certain variables, the right-hand side of Eq. (\ref{eq:MAPred}) can be written as
\begin{equation}
\label{eq:marglikelihood}
p(\textbf{w}\vert N) = \int p(\textbf{w}\vert\boldsymbol{\theta},N)p(\boldsymbol{\theta}\vert N)d^M\boldsymbol{\theta}
\end{equation}
where the marginalised variables are the parameters $\boldsymbol{\theta}$ of the expectation model. The first term in the integral, p(\textbf{w}$\vert\boldsymbol{\theta}$,N), is the likelihood function which describes the probability of the observed image pixel values \textbf{w} for particular values of the parameters $\boldsymbol{\theta}$ of a model with N atomic columns. Therefore, it is an explicit function of the parameters $\boldsymbol{\theta}$. In essence, the likelihood function is a measure of the goodness of fit of the model with the experimental measurements or image pixel values. The other term in the integral, p($\boldsymbol{\theta}\vert$N), is the prior density of the parameters $\boldsymbol{\theta}$ for a model with N columns. In practice, evaluation of the MAP probability rule can be reduced to calculating the marginal likelihood p(\textbf{w}$\vert$N) described by Eq. (\ref{eq:marglikelihood}) and determining for what number of columns N it is maximised. In order to do so, explicit expressions for the likelihood function p(\textbf{w}$\vert\boldsymbol{\theta}$,N) and the prior density p($\boldsymbol{\theta}\vert$N) are required. \\
\indent An expression for the likelihood function p(\textbf{w}$\vert\boldsymbol{\theta}$,N) can be derived by taking into account knowledge about the statistical properties of the errors in the experimental measurements. Since a STEM image of K$\times$L pixels is formed by counting electrons scattered to the detector, the pixel values are inevitably subject to Poisson noise \cite{haight(1967)} causing each observed image pixel value w$_{kl}$ at position (x$_k$,y$_l$) to be Poisson distributed \cite{mood(1974),vandenbos(2001)}. For an increasing expectation value f$_{kl}$($\boldsymbol{\theta}$) of w$_{kl}$, the Poisson distribution tends to be a normal distribution with mean $\mu_{kl}$ = f$_{kl}$($\boldsymbol{\theta}$) and standard deviation $\sigma_{kl}$ = [f$_{kl}$($\boldsymbol{\theta}$)]$^{1/2}$ \cite{papoulis(2002)}. Under the assumption that the pixel values are statistically independent, the likelihood function can be expressed as
\begin{equation}
\label{eq:normalgen}
p(\textbf{w}\vert\boldsymbol{\theta},N) = \frac{e^{-\chi^2(\boldsymbol{\theta})/2}}{\prod_{k=1}^K\prod_{l=1}^L[2\pi \sigma_{kl}^2]^{1/2}},
\end{equation}
where 
\begin{equation}
\label{eq:chigen}
\chi^2(\boldsymbol{\theta}) = \sum_{k=1}^K\sum_{l=1}^L\frac{[w_{kl}-\mu_{kl}]^2}{\sigma_{kl}^2}.
\end{equation}
For simplicity, it can be assumed that $\sigma_{kl} \approx [w_{kl}$]$^{1/2}$, so that $\sigma_{kl}$ is independent of the parameters $\boldsymbol{\theta}$. The likelihood function then becomes
\begin{equation}
\label{eq:normal}
p(\textbf{w}\vert\boldsymbol{\theta},N) = \frac{e^{-\chi^2(\boldsymbol{\theta})/2}}{\prod_{k=1}^K\prod_{l=1}^L[2\pi w_{kl}]^{1/2}},
\end{equation}
where
\begin{equation}
\label{eq:chi}
\chi^2(\boldsymbol{\theta}) = \sum_{k=1}^K\sum_{l=1}^L\frac{[w_{kl}-f_{kl}(\boldsymbol{\theta})]^2}{w_{kl}}
\end{equation}
is a weighted sum-of-squared-residuals misfit between the data and the parametric model. \\
\indent Different prior density functions p($\boldsymbol{\theta}\vert$N) can be constructed reflecting different types of prior knowledge. In this paper, p($\boldsymbol{\theta}\vert$N) is expressed as a product of uniform distributions over a predefined range for each parameter $\theta_m$:
\begin{equation}
\label{eq:prior1}
p(\boldsymbol{\theta}\vert N) = \begin{cases}
\prod_{m=1}^M\frac{1}{\theta_{m_{max}}-\theta_{m_{min}}} & \text{for} \; m=1, ..., M\text{:} \; \theta_{m_{min}} \leqslant \theta_m \leqslant \theta_{m_{max}}\\
0 & \text{otherwise}
\end{cases}
\end{equation}
where the subscripts \emph{max} and \emph{min} refer to a predefined maximum and minimum value, respectively. The choice for this form of prior simplifies the subsequent algebra significantly. More importantly, by using this uniform prior combined with a conservative choice of the predefined parameter ranges, the amount of prior knowledge that is introduced can be kept minimal, thus avoiding biased results due to the incorporation of possibly invalid prior knowledge. Moreover, this form of prior can also be used in a flexible way since the predefined parameter ranges can easily be adapted depending on the available prior knowledge. \\
\indent Finally, it can be shown that by substituting Eq. (\ref{eq:normal}) and Eq. (\ref{eq:prior1}) in Eq. (\ref{eq:marglikelihood}) an approximate analytical result for the posterior probability of the presence of N atomic columns  in the image, given the observed image pixel values $\textbf{w}$, is available. For the model described by Eq. (\ref{eq:model}), this expression is given by
\begin{equation}
\label{eq:MAPfinal}
p(N\vert\textbf{w}) \propto \frac{N!(4\pi)^{2N}e^{-\chi_{min}^2/2}[det(\nabla\nabla\chi^2)]^{-1/2}}{[(\beta_{x_{max}}-\beta_{x_{min}})(\beta_{y_{max}}-\beta_{y_{min}})(\eta_{max}-\eta_{min})(\rho_{max}-\rho_{min})]^N}
\end{equation}
in which $\chi^2_{min}$ = $\chi^2$($\hat{\boldsymbol{\theta}}$), with $\hat{\boldsymbol{\theta}}$ the parameter vector that maximizes the likelihood function described by Eq. (\ref{eq:normal}). The term det($\nabla\nabla\chi^2$) = det($\frac{\partial^2\chi^2(\boldsymbol{\theta})}{\partial\boldsymbol{\theta}\partial\boldsymbol{\theta}^T}\big|_{\boldsymbol{\theta}=\hat{\boldsymbol{\theta}}}$) represents the determinant of the Hessian matrix of $\chi^2(\boldsymbol{\theta})$ evaluated at $\hat{\boldsymbol{\theta}}$. It should be noted that the expression in Eq. (\ref{eq:MAPfinal}) is only valid if $\hat{\boldsymbol{\theta}}$ lies well within the support of the prior density function. In \ref{appendixA}, a more detailed explanation of its derivation and underlying assumptions is presented. The importance of Eq. (\ref{eq:MAPfinal}) is that it allows one to compute the posterior probability of a certain number of atomic columns present in a HAADF STEM image, for a model where the columns are assumed to be Gaussian shaped and to have different widths. The MAP probability rule compares the calculated posterior probabilities and selects the number of columns N with the highest probability. Similar expressions can be derived for other types of models. 
\begin{figure*}[ht]
\centering
\includegraphics{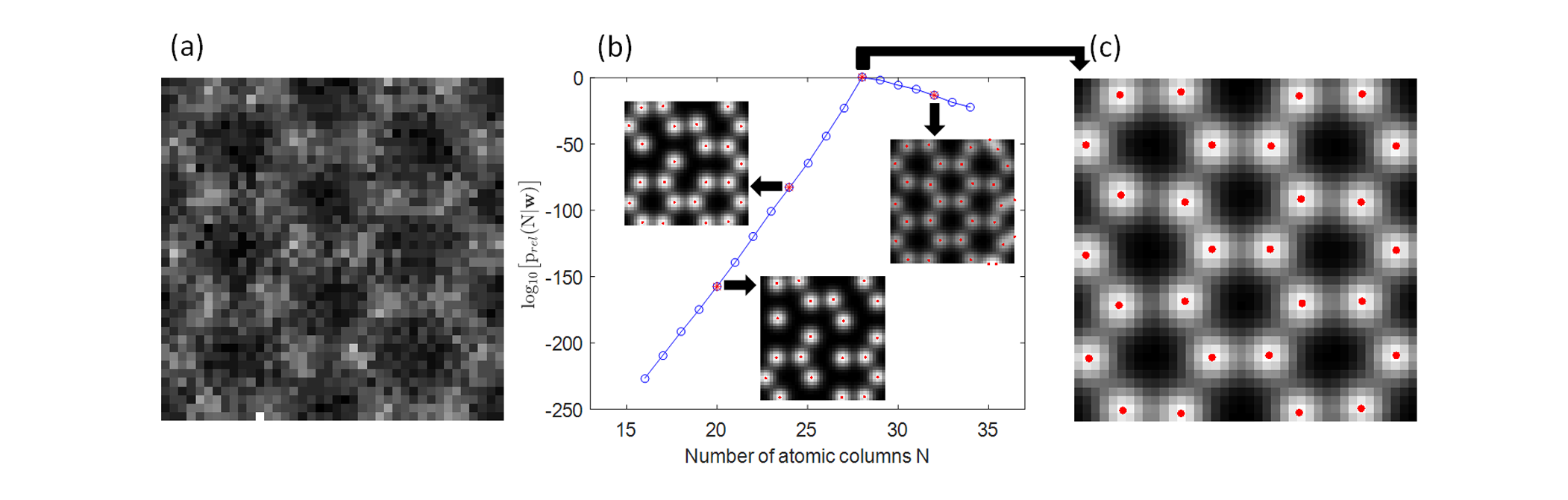}
\caption{(a) Simulated STEM image of graphene disturbed by Poisson noise with an incoming electron dose of 3$\cdot$10$^5$ e$^-$/$\AA^2$. (b) MAP probability rule evaluated for the data shown in (a). (c) Most probable parametric model of the data in (a) as indicated by the MAP probability rule in (b).}
\label{fig:1}
\end{figure*}
For example, for a model where the atomic columns are assumed to be Gaussian shaped and to have equal widths \cite{debacker(2016)}, the expectation model f$_{kl}$($\boldsymbol{\theta}$) of pixel (k,l) at position (x$_k$,y$_l$) is given by
\begin{equation}
\label{eq:model2}
f_{kl}(\boldsymbol{\theta}) =\zeta+\sum_{n=1}^{N}\eta_n exp\bigg(-\frac{(x_k-\beta_{x_n})^2+(y_l-\beta_{y_n})^2}{2\rho^2}\bigg)
\end{equation}
where the unknown parameter vector can be written as
\begin{equation}
\label{eq:parametervector2}
\boldsymbol{\theta}=(\beta_{x_1},...,\beta_{x_N},\beta_{y_1},...,\beta_{y_N},\eta_1,...,\eta_N,\rho,\zeta)^T
\end{equation}
containing M = 3N+2 parameters. For such a model, the posterior probability becomes
\begin{equation}
\label{eq:MAPfinalequal}
p(N\vert\textbf{w}) \propto \frac{N!(4\pi)^{1.5N}e^{-\chi_{min}^2/2}[det(\nabla\nabla\chi^2)]^{-1/2}}{[(\beta_{x_{max}}-\beta_{x_{min}})(\beta_{y_{max}}-\beta_{y_{min}})(\eta_{max}-\eta_{min})]^N}.
\end{equation}
As another example, for an expectation model given by
\begin{equation}
\label{eq:model3}
f_{kl}(\boldsymbol{\theta}) =\zeta+\sum_{n=1}^{N}\eta exp\bigg(-\frac{(x_k-\beta_{x_n})^2+(y_l-\beta_{y_n})^2}{2\rho^2}\bigg),
\end{equation}
where the columns are assumed to be Gaussian shaped with equal widths and equal heights, with unknown parameter vector
\begin{equation}
\label{eq:parametervector3}
\boldsymbol{\theta}=(\beta_{x_1},...,\beta_{x_N},\beta_{y_1},...,\beta_{y_N},\eta,\rho,\zeta)^T
\end{equation}
containing M = 2N+3 parameters, the posterior probability becomes
\begin{equation}
\label{eq:MAPfinalequal2}
p(N\vert\textbf{w}) \propto \frac{N!(4\pi)^{N}e^{-\chi_{min}^2/2}[det(\nabla\nabla\chi^2)]^{-1/2}}{[(\beta_{x_{max}}-\beta_{x_{min}})(\beta_{y_{max}}-\beta_{y_{min}})]^N}.
\end{equation}
Also for models where other shapes of atomic columns are assumed, such as Lorentzian shapes or a mixture of Lorentzian and Gaussian shapes \cite{verbeeck(2012)}, an expression for the posterior probability as a function of the number of atomic columns can be derived in a similar way. It should be noted that the analytical expression for p(N$\vert\textbf{w}$) was derived under the assumption that the Poisson distribution that governs the image pixel values can be approximated by a normal distribution. The accuracy of this approximation, and therefore the accuracy of the expressions given by Eqs. (\ref{eq:MAPfinal}), (\ref{eq:MAPfinalequal}) and (\ref{eq:MAPfinalequal2}), increases with an increasing amount of detected electrons. The expressions will be most accurate if all image pixel values fully satisfy the normality assumption, but have shown to be robust to small violations of this assumption. Therefore, the MAP probability rule is an adequate method allowing to decide the number of atomic columns present in a STEM image.

\subsection{Algorithm}
As stated above, the MAP probability rule selects the most probable number of atomic columns by comparing posterior probabilities as a function of N, i.e. for different numbers of Gaussian peaks in the model describing the atomic columns in the image, starting from an initial model containing N$_0$ peaks up to and including a model containing a value of N$_{max}$ peaks. The parameters $\boldsymbol{\theta}$ of the initial model are optimized by minimizing the weighted sum-of-squared-residuals misfit $\chi^2(\boldsymbol{\theta})$, given by Eq. (\ref{eq:chi}), subject to the constraint that $\boldsymbol{\theta}$ should belong to the support of the prior density function described by Eq. (\ref{eq:prior1}). Next, an extra peak is added to the initial configuration, so a model is constructed containing N$_0$+1 peaks. Again, the parameters of this model are optimized by minimizing $\chi^2(\boldsymbol{\theta})$ subject to the constraint that $\boldsymbol{\theta}$ belongs to the support of the prior density function. To avoid ending up in a local minimum, many different starting positions for the extra added peak need to be tested. To optimize the parameters associated with the other peaks, the estimated parameter values of the previous optimization, in this case of a model with N$_0$ peaks, are used as starting values. Next, another peak is added, in the same way as described above, in order to obtain the optimal parameter values of a model containing N$_0$+2 peaks. This procedure continues until the parameters of a model with N$_{max}$ peaks are optimized. In order to determine the most probable number of atomic columns present in a STEM image, the posterior probability of  N = N$_0$,...,N$_{max}$ columns is computed relatively to the posterior probability of N$_{max}$ columns as follows
\begin{equation}
\label{eq:MAPrel}
p_{rel}(N\vert\textbf{w}) = \frac{p(N\vert\textbf{w})/p(N_{max}\vert\textbf{w})}{max_N\bigg[p(N\vert\textbf{w})/p(N_{max}\vert\textbf{w})\bigg]}
\end{equation}
where the denominator is a normalization constant so that the maximum value of p$_{rel}$(N$\vert\textbf{w}$) corresponds to one. 
\begin{table}[b]\footnotesize
\centering
\begin{tabular}{ l l l}
\hline
Parameter & Symbol & Value \\
\hline
Acceleration voltage & V$_0$ (kV) & 80 \\
Defocus & $\epsilon$ ($\AA$) & -20.0 \\
Spherical aberration & C$_s$ (mm) & 0.0037 \\
Spherical aberration of 5th order & C$_5$ (mm) & 0 \\
Semiconvergence angle & $\alpha$ (mrad) & 24.8 \\
Detector inner radius & $\beta_1$ (mrad) & 26 \\
Detector outer radius & $\beta_2$ (mrad) & 50 \\
Pixel size & $\Delta$x=$\Delta$y ($\AA$) & 0.20 \\
FWHM of the source image & FWHM ($\AA$) & 0.7 \\
\hline
\end{tabular}
\caption{Microscope parameter values for simulation of a STEM image of graphene using MULTEM.}
\label{tab:graphene}
\end{table}
The most probable number of atomic columns is then given by the value N that maximizes Eq. (\ref{eq:MAPrel}). Direct visualization of the probability of the number of columns in the image is possible by plotting p$_{rel}$(N$\vert\textbf{w}$) on a logarithmic scale as a function of N resulting in a relative probability curve. This procedure has been illustrated in Fig.~\ref{fig:1} where the MAP probability rule has been applied to detect the C atoms of graphene from a Poisson disturbed simulated STEM image shown in Fig.~\ref{fig:1}(a), using a parametric model described by a superposition of Gaussian peaks with equal widths and equal intensities given by Eq. (\ref{eq:model3}). The image has been simulated using the MULTEM software \cite{lobato(2015),lobato(2016)} and the simulation parameters are listed in Table \ref{tab:graphene}. Fig.~\ref{fig:1}(b) shows the relative probability curve as a function of the number of columns N calculated by the decimal logarithm of Eq. (\ref{eq:MAPrel}). For certain values of N, the corresponding optimized models are shown. The most probable structure found by the MAP probability rule applied to the image shown in Fig.~\ref{fig:1}(a) is shown in Fig.~\ref{fig:1}(c), corresponding to the expected hexagonal lattice of graphene. The procedure described in this section has also been applied successfully to determine the structure of SrTiO$_3$, a Au nanorod and ultrasmall Ge clusters from both simulated and experimental STEM images exhibiting low CNR \cite{fatermans(2018)}. The approach has been implemented in the freely available StatSTEM software \cite{debacker(2016)}.

\section{Atom detectability}\label{quality}
Besides detecting atomic columns from HAADF STEM images, the MAP probability rule also offers a way to evaluate the relation between image quality measures and atom detectability. A common HAADF STEM image quality measure is the SNR. In general, one expects that the detectability of the atomic columns in the image will increase with increasing SNR. Note that, as local fluctuations of the background or differences in column thickness or composition can occur, each column can possess a different SNR value. The SNR of a column located at pixel (k,l) can be defined as
\begin{equation}
\label{eq:snr}
SNR = \frac{\lambda_{kl}}{\sigma_{kl}}
\end{equation}
where $\lambda_{kl}$ and $\sigma_{kl}$ denote the expected value of the observed image pixel value w$_{kl}$ at position (x$_k$,y$_l$) and its standard deviation, respectively. As w$_{kl}$ is Poisson distributed, $\sigma_{kl}$ = [$\lambda_{kl}]^{1/2}$. When the background in the image can be considered to be constant, Eq. (\ref{eq:snr}) can be written as
\begin{equation}
\label{eq:snr3}
SNR = \frac{\eta+\zeta}{[\eta+\zeta]^{1/2}}
\end{equation}
where $\eta$ and $\zeta$ denote the height of the atomic column and the background of the image, respectively, for a parametric model based on a superposition of Gaussian peaks, such as in Eqs. (\ref{eq:model}), (\ref{eq:model2}) or (\ref{eq:model3}). An alternative measure to describe the quality of an image is given by the CNR \cite{welvaert(2013)}. When a parametric model is used to describe the background and atomic columns in a HAADF STEM image, the CNR of a column can be defined as
\begin{equation}
\label{eq:cnr2}
CNR = \frac{\eta}{[\eta+\zeta]^{1/2}}.
\end{equation}
Note that the definition of CNR in Eq. (\ref{eq:cnr2}) is closely related to SNR in Eq. (\ref{eq:snr3}), but in case of CNR the background $\zeta$ is subtracted before taking the ratio. As such, the CNR also takes image contrast into account.
\begin{table}[b]\footnotesize
\centering
\begin{tabular}{ l l l}
\hline
Parameter & Symbol & Value \\
\hline
Acceleration voltage & V$_0$ (kV) & 120 \\
Defocus & $\epsilon$ ($\AA$) & 0 \\
Spherical aberration & C$_s$ (mm) & 0.001 \\
Spherical aberration of 5th order & C$_5$ (mm) & 0 \\
Semiconvergence angle & $\alpha$ (mrad) & 21.3 \\
Detector inner radius & $\beta_1$ (mrad) & 28 \\
Detector outer radius & $\beta_2$ (mrad) & 172 \\
FWHM of the source image & FWHM ($\AA$) & 0.7 \\
\hline
\end{tabular}
\caption{Microscope parameter values for simulation of a set of HAADF STEM images of 12.5 $\AA$ by 12.5 $\AA$ using MULTEM.}
\label{tab:parameters}
\end{table}

\subsection{Comparison of SNR to CNR}
\begin{figure*}[ht]
\centering
\includegraphics{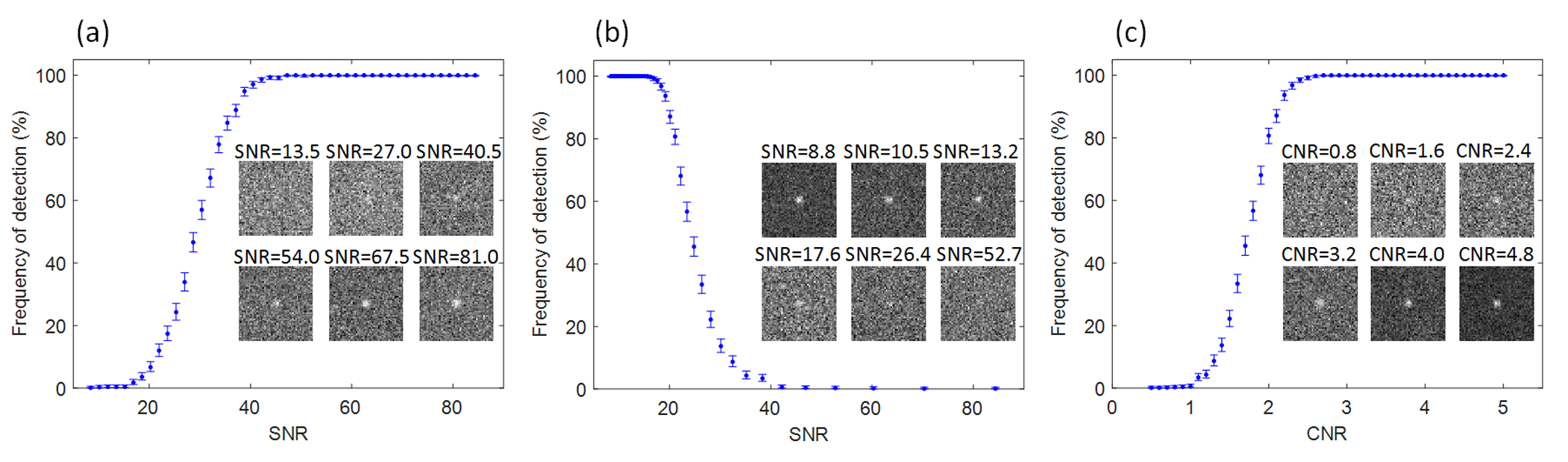}
\caption{(a) Observed detection rate of a Au atom by the MAP probability rule from simulated HAADF STEM images of 12.5 $\AA$ by 12.5 $\AA$ with a pixel size of 0.25 $\AA$ as a function of SNR where the SNR has been altered by changing the incoming electron dose ranging from 10$^3$ e$^-$/$\AA^2$ to 10$^5$ e$^-$/$\AA^2$ and (b) by changing the constant background for a fixed incoming electron dose of 10$^4$ e$^-$/$\AA^2$. (c) Detection rate from the same set of images as in (b) as a function of CNR. The insets show simulated images disturbed by Poisson noise for different values of SNR and CNR.}
\label{fig:2}
\end{figure*}
In this section, the MAP probability rule has been used as a tool to compare SNR and CNR in relation with atom detectability. First, SNR is investigated. For this, a set of simulated HAADF STEM images of 12.5 $\AA$ by 12.5 $\AA$ with a pixel size of 0.25 $\AA$ has been simulated of an individual Au atom using MULTEM to which an arbitrary constant background has been added accounting for the contribution of electrons scattered by an amorphous substrate. The remaining simulation parameters are listed in Table~\ref{tab:parameters}. Each simulated image has been generated 1000 times containing random Poisson noise. The SNR of the atom in the image has been altered by changing the incoming electron dose ranging from 10$^3$ e$^-$/$\AA^2$ to 10$^5$ e$^-$/$\AA^2$ resulting in a higher SNR value for a higher incoming electron dose. For detecting the Au atom from the noise disturbed images by the proposed MAP probability rule, a model assuming the image of the atom to be Gaussian shaped has been used, where a constant background $\zeta$ and width $\rho$, height $\eta$ and x- and y-coordinate $\beta_x$ and $\beta_y$ of the atom need to be estimated. The prior density p($\boldsymbol{\theta}\vert$N) has been chosen according to Eq. (\ref{eq:prior1}), where the parameters $\zeta$ and $\eta$ range from 0 up to the maximum pixel intensity in the simulated image, whereas the parameters $\rho$, $\beta_x$ and $\beta_y$ range according to the field of view of the image, i.e. from 0 $\AA$ up to 12.5 $\AA$. Fig~\ref{fig:2}(a) shows the observed detection rates of the MAP probability rule as a function of SNR. The error bars show the 95 \% confidence Wilson score intervals of a binomial distribution \cite{wilson(1927)}. The inset shows simulated STEM images containing Poisson noise for different SNR values.
\begin{figure*}[ht]
\centering
\includegraphics{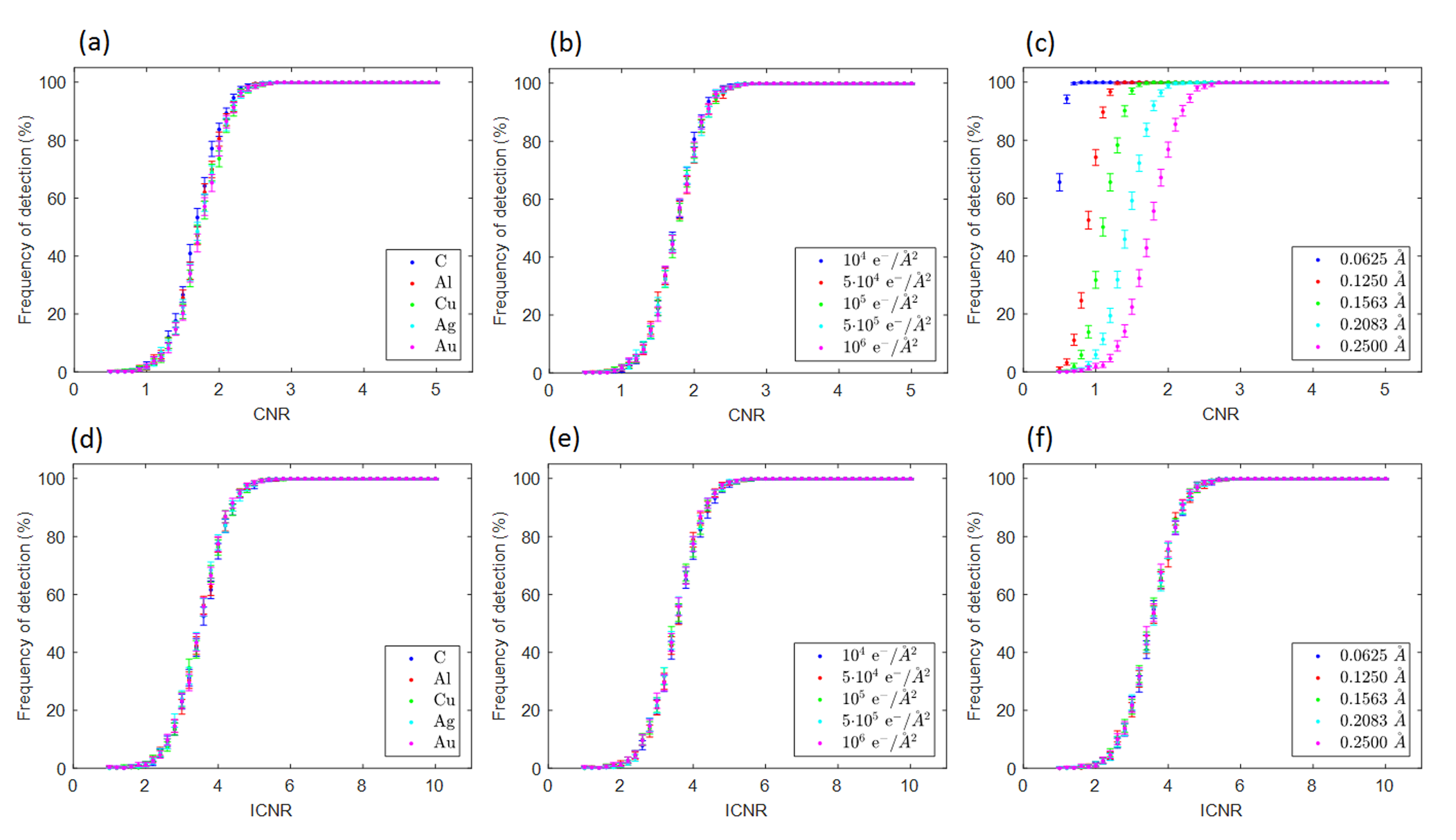}
\caption{Observed detection rate of an individual atom by the MAP probability rule from simulated HAADF STEM images of 12.5 $\AA$ by 12.5 $\AA$ with a pixel size of 0.25 $\AA$ for varying atom types with an incoming electron dose of 10$^6$ e$^-$/$\AA^2$ as a function of (a) CNR and (d) ICNR. Detection rate of a Au atom from simulated HAADF STEM images of 12.5 $\AA$ by 12.5 $\AA$ with a pixel size of 0.25 $\AA$ for varying incoming electron dose as a function of (b) CNR and (d) ICNR and from images with an incoming electron dose of 10$^5$ e$^-$/$\AA^2$ for varying pixel size as a function of (c) CNR and (f) ICNR.}
\label{fig:3}
\end{figure*}
Alternatively, the SNR value of the Au atom in the simulated image can be altered by changing the added constant background independently of the incoming electron dose, simulating the effect of the atom to be positioned on a substrate with varying thickness. By altering the SNR by this procedure for a fixed incoming electron dose of 10$^4$ e$^-$/$\AA^2$ and by using the same approach for the MAP probability rule as for Fig.~\ref{fig:2}(a), Fig.~\ref{fig:2}(b) shows that the detection rate decreases with increasing SNR value, as opposed to Fig.~\ref{fig:2}(a). Apparently, a high SNR value does not guarantee high atom detectability, as is also visually perceived by the inset of Fig.~\ref{fig:2}(b) showing Poisson disturbed simulated STEM images for different SNR values. The reason for this behaviour lies in the fact that the SNR measure given by Eq. (\ref{eq:snr3}) only considers the total sum of the height of the atomic column $\eta$ and the background of the image $\zeta$. Therefore, SNR does not take image contrast into account and it is possible that the SNR of an image is high while contrast is low. Contrast, however, also affects the visual perception and detectability of objects in an image. The CNR measure given by Eq. (\ref{eq:cnr2}) provides information about image contrast as the background $\zeta$ is explicitly subtracted. Therefore, CNR relates better to atom detectability than SNR. This is confirmed by Fig.~\ref{fig:2}(c) where the observed detection rate is shown as a function of the CNR for the same set of images as of Fig.~\ref{fig:2}(b). From this, it is seen that the detection rate increases with increasing CNR. It is noted that the same behaviour is observed for the set of images of Fig.~\ref{fig:2}(a).

\subsection{Integrated CNR (ICNR)}
As followed from the previous section, CNR is a more intuitive image quality measure than SNR when it comes to detecting atoms from HAADF STEM images. In this section, the relation between the CNR measure and atom detectability is investigated for variations of atom type, incoming electron dose and image pixel size. First, the detection rate of the MAP probability rule for different types of individual atoms as a function of CNR is shown in Fig.~\ref{fig:3}(a) for simulated HAADF STEM images of 12.5 $\AA$ by 12.5 $\AA$ with a pixel size of 0.25 $\AA$, using MULTEM with an incoming electron dose of 10$^6$ e$^-$/$\AA^2$, where each simulated image has been generated 1000 times containing Poisson noise. The remaining simulation parameters can be found in Table~\ref{tab:parameters}. The CNR of the atom in the image has been altered by adding a constant background. For detecting the atom from the noise disturbed images by the MAP probability rule, the same approach as in Fig.~\ref{fig:2} has been followed with an equivalent choice for the prior density p($\boldsymbol{\theta}\vert$N). It follows from Fig.~\ref{fig:3}(a) that the relation between CNR and detection rate depends hardly, yet slightly, on the atom type. Next, the relation between CNR and detection rate to varying incoming electron dose is investigated. For this, images of 12.5 $\AA$ by 12.5 $\AA$ with a pixel size of 0.25 $\AA$ of an individual Au atom have been simulated with different incoming electron doses. Fig.~\ref{fig:3}(b) shows that the relation between CNR and the atom detection rate is rather robust to a different incoming electron dose. On the other hand, a different pixel size influences the detection rate as a function of CNR significantly, as shown in Fig.~\ref{fig:3}(c) for images of a Au atom with an incoming electron dose of 10$^5$ e$^-$/$\AA^2$ and varying pixel size. This is due to the fact that, for a fixed incoming electron dose/$\AA^2$, an increased pixel size leads to an increased electron dose/pixel resulting into a higher value for the background $\zeta$ and height $\eta$ of the Au atom, as both $\zeta$ and $\eta$ scale with electron dose/pixel, and hence into a higher CNR given by Eq. (\ref{eq:cnr2}) and vice versa. For this reason, the integrated CNR (ICNR) is proposed whose relationship with atom detectability is independent of atom type, incoming electron dose and, in particular, the pixel size of the image. The ICNR of an atomic column in HAADF STEM images is defined as the ratio of the total intensity of electrons scattered by the column, the so-called scattering cross section \cite{debacker(2016)},\cite{vanaert(2009)},\cite{vanaert(2013)},\cite{e(2013)}, to the square root of the sum of the scattering cross section and the integrated background under the column. When a parametric model based on a superposition of Gaussian peaks is used to describe the background and atomic columns, the ICNR of a column can be calculated as
\begin{equation}
\label{eq:ICNR}
ICNR = \frac{2\pi\eta\rho^2}{[2\pi\eta\rho^2+\pi(3\rho)^2\zeta]^{1/2}},
\end{equation}
where $\rho$ denotes the estimated width of the atomic column. In this work, ICNR values are calculated by expressing $\rho$ in units of pixels. In order to estimate the integrated background under the column, the area under the column has been considered to be a circle with a radius of 3$\rho$, since 99.46 \% of the volume under the Gaussian peak describing the column is contained within this distance. It should be noted that Eq. (\ref{eq:ICNR}) is only valid for individual atomic columns that are well separated in the image. For denser crystallographic structures, the contrast of a column depends, not only on the height of the column $\eta$ and on the background in the image $\zeta$, but also on the heights of and the distances from the surrounding columns. For the investigation of the relation between ICNR and atom detectability to varying atom type, incoming electron dose and pixel size, the same procedure for the simulated images has been followed as for the investigation of CNR. It is shown in Fig.~\ref{fig:3}(d) and ~\ref{fig:3}(e) that the detection rate does not change with atom type or incoming electron dose, respectively, as long as the ICNR value remains unchanged. The functional relationship between detection rate and ICNR is also independent of pixel size, as shown in Fig.~\ref{fig:3}(f), as opposed to the relationship between detection rate and CNR, as shown in Fig.~\ref{fig:3}(c). The results indicate that ICNR is a more robust measure for atom detectability from high-resolution HAADF STEM images than CNR.

\section{Relation to model selection}\label{modelselection}
In this section, it is explained how the MAP probability rule is related to the concept of model selection, where one aims to select the best model from a set of candidate models given experimental data. For this, the working principle of model selection is described in detail. Hereby, different model-order selection criteria are introduced in order to compare their performances to detect the correct number of atomic columns from HAADF STEM images to that of the MAP probability rule. 

\subsection{Model complexity}
The heart of model selection consists of selecting one model from a set of competing models that represents most closely the underlying process that generated the experimental data. For this purpose, a criterion measuring how well the model fits the data is required. Such a criterion of goodness of fit quantifies the descriptive adequacy of a model, which is possible by, for example, a maximum likelihood evaluation. However, a model-selection criterion based solely on goodness of fit automatically selects the model which fits best to the data. This is undesired, since model fit can be easily improved by increased model complexity, referring to the flexibility of a model to fit the observed data. In this way, a model might be selected without necessarily bearing any interpretable relationship with the underlying data-generating process. For this reason, typical model selection methods take into account both the goodness of fit and the complexity of the models under investigation \cite{myung(2000)}. \\
\indent There are at least three important factors that contribute to the complexity of a model \cite{myung(1997)}. The first one is the number of parameters. In general, a model with many parameters describes data better than a model with few parameters due to its higher flexibility, and hence complexity. Next, model complexity is also related to functional form, which is described as the way in which the parameters are combined in the model. A model with a more complex functional form is able to describe a wider range of data and can be considered to be more flexible than a model with a less complex functional form. The last dimension of model complexity is covered by the extension of the parameter space. A model of which the parameters can fluctuate over a wide range of values can describe a wider range of data. Therefore, such a model is considered to be more complex than a model of which the parameters can fluctuate over only a small range. All of these three aspects can significantly and independently influence model fit. \\
\indent Typically, model-selection criteria are written as twice the negative log likelihood, accounting for the goodness of fit, plus a penalty term that accounts for model complexity \cite{claeskens(2008)}:
\begin{equation}
\label{eq:criteria}
-2log(\hat{L}) + 2C.
\end{equation}
Different criteria have been proposed in the literature and describe the complexity of the model in a different way, often taking only one dimension of model complexity into account, i.e. the number of parameters \cite{stoica(2004)}. The Akaike Information Criterion (AIC) \cite{akaike(1974)} can be written as
\begin{equation}
\label{eq:AIC}
AIC = -2log(\hat{L}) + 2M
\end{equation}
with M the number of parameters in the model. A very similar criterion is the Generalized Information Criterion (GIC) \cite{broersen(1993)}, where the contribution of the penalty term can be modified by the parameter d:
\begin{equation}
\label{eq:GIC}
GIC = -2log(\hat{L}) + dM.
\end{equation}
A clear guideline on how to choose the value of d is lacking. Different choices of d = 3 \cite{broersen(1996)} and d = 4 \cite{stoica(2004)} are reported in the literature, which are referred to, in what follows, as GIC3 and GIC4, respectively. These choices cause the GIC to penalise more heavily for the complexity of the model as compared to the AIC. Another common criterion is the Bayesian Information Criterion (BIC) \cite{schwarz(1978)}: 
\begin{equation}
\label{eq:BIC}
BIC = -2log(\hat{L}) + Mlog(W)
\end{equation}
with W the sample size (i.e. the number of datapoints).  For a STEM image, W is equal to K$\times$L pixels. As opposed to the AIC and GIC, the penalty term is dependent on the sample size. When W $>$ 8, the BIC accounts more for the complexity of the model than the AIC. An alternative criterion is the Hannan-Quinn Information Criterion (HQC) \cite{hannan(1979)} which replaces the log(W) factor in the BIC by the slower diverging quantity log[log(W)]:
\begin{equation}
\label{eq:HQC}
HQC = -2log(\hat{L}) + Mlog[log(W)].\\
\end{equation}
\begin{figure*}[ht]
\centering
\includegraphics{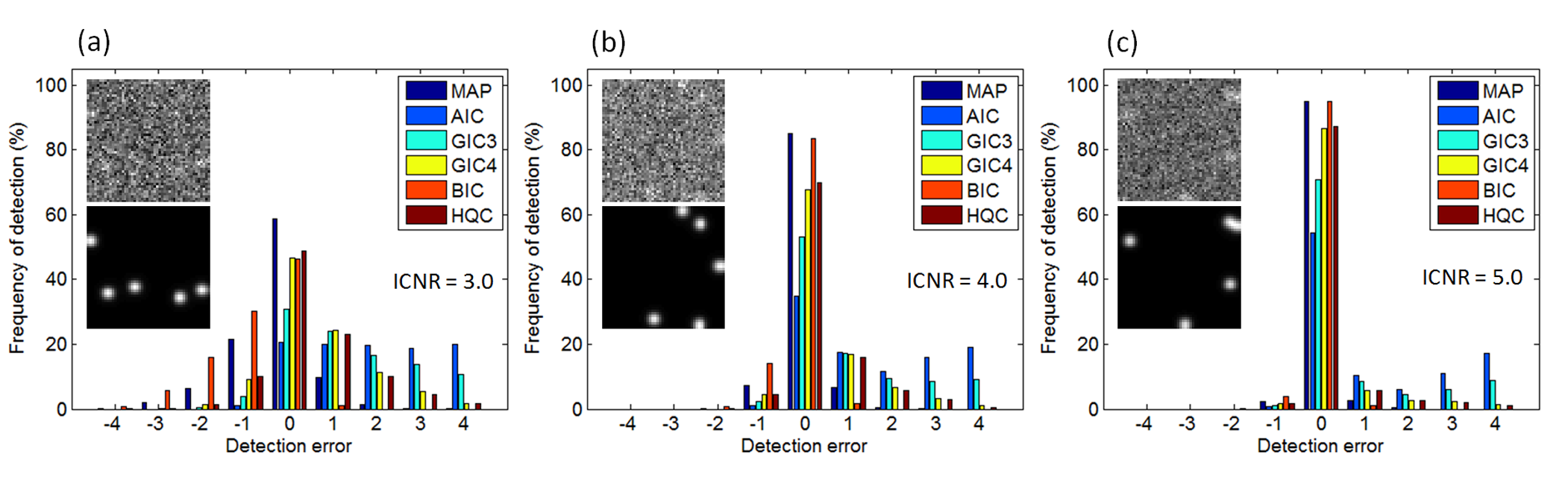}
\caption{Average frequency of various selected orders for different model selection-criteria for detecting the number of Au atoms from a set of simulated HAADF STEM images of 12.5 $\AA$ by 12.5 $\AA$ with a pixel size of 0.25 $\AA$ and with (a) an ICNR value of 3.0, (b) an ICNR value of 4.0, and (c) an ICNR value of 5.0. The insets show randomly generated simulated images of 5 atoms disturbed by Poisson noise with the noise-free images as references.}
\label{fig:4}
\end{figure*}
\indent Thus, a model selection method performs a tradeoff between high goodness of fit and low model complexity. In essence, the MAP probability rule is also a model selection method where the posterior probability of the presence of N atomic columns in the STEM image can be written in general terms as
\begin{equation}
\label{eq:generalselection}
p(N\vert\textbf{w}) \propto \frac{\text{goodness of fit}}{\text{model complexity}}
\end{equation}
where the numerator, goodness of fit, is the likelihood given by Eq. (\ref{eq:normal}) evaluated at $\hat{\boldsymbol{\theta}}$. The denominator describes the model complexity and is given by
\begin{equation}
\label{eq:complex}
\text{model complexity} = \frac{[det(\nabla\nabla\chi^2)]^{1/2}}{N!(4\pi)^{M/2}p(\boldsymbol{\theta}\vert N)}
\end{equation}
when the approximations of the analytical expression of p(N$\vert$\textbf{w}) are valid. A more detailed explanation of Eq. (\ref{eq:complex}) can be found in \ref{appendixB}. The term $\nabla\nabla\chi^2$ is a M$\times$M dimensional matrix and therefore it depends on the number of parameters. In addition, it explicitly contains the expectation model f$_{kl}$($\boldsymbol{\theta}$), describing the intensity of pixel (k,l) at position (x$_k$,y$_l$), as follows from Eq. (\ref{eq:chi}). Therefore, the term $\nabla\nabla\chi^2$ in Eq. (\ref{eq:complex}) takes into account two dimensions of model complexity, which are the number of parameters on the one hand and the functional form of the model on the other hand. The third dimension of model complexity, which is the extension of the parameter space, is described by the prior density p($\boldsymbol{\theta}\vert$N). By choosing p($\boldsymbol{\theta}\vert$N) as a product of uniform distributions for each parameter $\theta_m$ individually, as given by Eq. (\ref{eq:prior1}), large ranges for the possible values of $\theta_m$ correspond with a small value for p($\boldsymbol{\theta}\vert$N). As model complexity is inversely proportional to p($\boldsymbol{\theta}\vert$N), model complexity increases as the extension of the parameter space increases. As a result, the complexity term of the MAP probability rule depends on three dimensions of model complexity as opposed to the aforementioned AIC, GIC, BIC and HQC whose complexity terms only depend on the number of parameters M. By taking the logarithm of Eq. (\ref{eq:generalselection}) multiplied by -2 and taking into account the expression of model complexity of Eq. (\ref{eq:complex}), Eq. (\ref{eq:generalselection}) can be written in the same form as Eq. (\ref{eq:criteria}): 
\begin{equation}
\label{eq:MAPcriterion}
\begin{split}
-2log[p(N\vert\textbf{w})] &= -2log(\hat{L}) + log[det(\nabla\nabla\chi^2)] - 2log(N!)\\
 &- Mlog(4\pi) - 2log[p(\boldsymbol{\theta}\vert N)] + cst
\end{split}
\end{equation}
where the term cst refers to a constant coming from the proportionality of Eq. (\ref{eq:generalselection}). The second term at the right-hand side of Eq.~(\ref{eq:MAPcriterion}) can be written as
\begin{equation}
\label{eq:detchi}
\begin{split}
log[det(\nabla\nabla\chi^2)] &= log[det(W\cdot\frac{1}{W}\nabla\nabla\chi^2)] \\
&= Mlog(W) + log[det(\frac{1}{W}\nabla\nabla\chi^2)],
\end{split}
\end{equation}
where use is made of the fact that det(cA) = c$^M$det(A) for a scalar c and M$\times$M matrix A. As such, Eq. (\ref{eq:MAPcriterion}) can be rewritten as
\begin{equation}
\label{eq:MAPcriterion2}
\begin{split}
-2log[p(N\vert\textbf{w})] &= -2log(\hat{L}) + Mlog(W) + log[det(\frac{1}{W}\nabla\nabla\chi^2)]\\
 & - 2log(N!) - Mlog(4\pi) - 2log[p(\boldsymbol{\theta}\vert N)] + cst.
\end{split}
\end{equation}
Interestingly, the first two terms of Eq. (\ref{eq:MAPcriterion2}) correspond to the BIC described by Eq. (\ref{eq:BIC}). This indicates a relation between the MAP probability rule and the BIC, which is not surpising as both techniques are based on a Bayesian approach.

\subsection{Performance}\label{performance}
\indent In this section, the performance of the MAP probability rule in detecting atoms from STEM images is compared to the AIC, GIC, BIC and HQC, which were introduced in the previous section. For this, a set of 1000 images of Au atoms has been simulated by MULTEM with dimensions of 12.5 $\AA$ by 12.5 $\AA$ and a pixel size of 0.25 $\AA$. The remaining simulation parameters are listed in Table~\ref{tab:parameters}. Hereby, the number of Au atoms in an individual image is uniformly distributed between 1 and 5 atoms. The atoms are randomly positioned within the field of view of the image according to a uniform distribution and the incoming electron dose of the image can uniformly fluctuate between 5$\cdot$10$^3$ e$^-$/$\AA^2$ and 10$^5$ e$^-$/$\AA^2$, affecting the peak intensities of the atoms. Depending on the incoming electron dose, a constant background has been added to an individual image such that all 1000 simulated images have the same ICNR value. Hereby, each simulated image has been disturbed by Poisson noise. The performances of the model-selection criteria are evaluated by comparing the detected number of atoms in the image with the true number of atoms by calculating the average frequency of various selected orders \cite{stoica2(2004)}. For the analysis, a superposition of Gaussian peaks with equal widths and equal intensities has been used to model the Au atoms in the image. This model is described by Eq. (\ref{eq:model3}). Since the width $\rho$ has been considered to be a fixed value, the model consists of 2N+2 parameters: background $\zeta$, height $\eta$ and x- and y-coordinates $\beta_{x_n}$ and $\beta_{y_n}$. For the MAP probability rule, the prior density p($\boldsymbol{\theta}\vert$N) was chosen according to Eq. (\ref{eq:prior1}) where the background $\zeta$ ranges from 0 up to the maximum pixel intensity in the simulated image and where the distributions for the height $\eta$ and x- and y-coordinates $\beta_{x_n}$ and $\beta_{y_n}$ exactly correspond with the uniform distributions that were used to generate these parameters. The performances of the model-selection criteria are shown in Fig.~\ref{fig:4} for three different ICNR values. A positive value of the detection error refers to the detection of too many atoms, whereas a negative value refers to the detection of too few atoms. The insets show randomly generated simulated images of 5 atoms disturbed by Poisson noise with the noise-free images as references. From Fig.~\ref{fig:4}(a), it can be seen that the MAP probability rule outperforms the other criteria for an ICNR value of 3.0. The AIC, GIC3, GIC4 and HQC have a tendency to detect too many atoms, whereas the BIC often detects too few atoms. This behaviour is related to the different ways the model-selection criteria penalize the complexity of the model. From Fig.~\ref{fig:3}, showing that the detection rate increases when the ICNR of an image increases, it is expected that the frequency of detecting the correct number of atoms increases for increasing ICNR. This is confirmed in Fig.~\ref{fig:4}(b) and ~\ref{fig:4}(c) for an ICNR value of 4.0 and 5.0, respectively. Moreover, in these cases, the performances of the different criteria tend to become more equal.

\subsection{Relation between MAP and BIC}\label{BIC}
\begin{figure*}[ht]
\centering
\includegraphics{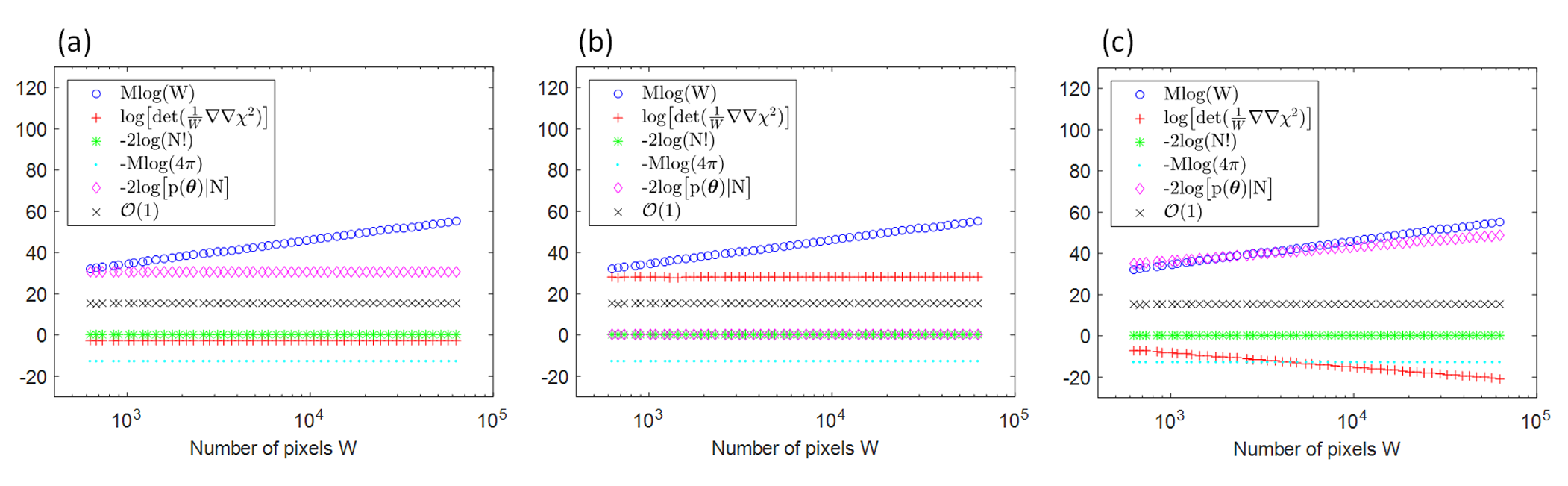}
\caption{Behaviour of the individual terms of the MAP probability rule in function of the number of pixels W for a set of simulated HAADF STEM images of 12.5 $\AA$ by 12.5 $\AA$ of a single Au atom with a pixel size ranging from 0.5 $\AA$ until 0.05 $\AA$ and an incoming electron dose of 625 e$^-$/pixel. For clarity, the term -2log($\hat{L}$) has not been visualized. Three different parametrizations of the model describing the STEM images have been used. In (a) the background $\zeta$ and height $\eta$ have been fitted in electron counts and the width $\rho$ and x- and y-coordinate $\beta_x$ and $\beta_y$ in $\AA$, in (b) the parameters have been normalized between 0 and 1 and in (c) $\zeta$ and $\eta$ have been fitted in electron counts and $\rho$, $\beta_x$ and $\beta_y$ in pixels.}
\label{fig:5}
\end{figure*}
As mentioned earlier, Eq. (\ref{eq:MAPcriterion2}) indicates a relation between the MAP probability rule and the BIC. In order to understand this relation, the behaviour of the terms in Eq. (\ref{eq:MAPcriterion2}) has been investigated as a function of the sample size by simulating a set of HAADF STEM images of 12.5 $\AA$ by 12.5 $\AA$ of a single Au atom. The STEM simulations have been obtained from MULTEM and the simulation parameters are listed in Table~\ref{tab:parameters}. In order to acquire a set of simulated images containing an increasing number of pixels W, the pixel size has been decreased starting from 0.5 $\AA$ up to and including 0.05 $\AA$. As such, 50 images have been simulated where for each image 20 random Poisson noise configurations have been applied. Hereby, the incoming electron dose has been set to 625 e$^-$/pixel to keep the amount of detected electrons per pixel constant, irrespective of the pixel size, so that the behaviour of the terms of the MAP probability rule is only dependent on the number of pixels W. Note that this means that the electron dose/$\AA^2$ increases as the pixel size decreases. For the analysis of the MAP probability rule, a model where the atom is assumed to be Gaussian shaped has been used consisting of M = 5 parameters: a constant background $\zeta$, width $\rho$, height $\eta$ and x- and y-coordinate $\beta_x$ and $\beta_y$. The prior density p($\boldsymbol{\theta}\vert$N) has been chosen according to Eq. (\ref{eq:prior1}) where the parameters $\zeta$ and $\eta$ range from 0 up to the maximum pixel intensity in the simulated image, whereas the parameters $\rho$, $\beta_x$ and $\beta_y$ range according to the field of view of the image, i.e. from 0 $\AA$ up to 12.5 $\AA$. The different terms contributing to the MAP probability rule given by Eq. (\ref{eq:MAPcriterion2}) are shown in Fig. ~\ref{fig:5}. The term -2log($\hat{L}$) depends on W and increases with an increasing number of pixels, but, for clarity, it has not been visualized since it would dominate the graphs as it has values up to three orders of magnitude larger than the other terms. It can be seen from Fig.~\ref{fig:5}(a) that the term Mlog(W) increases as the number of pixels W increases, as expected. The terms log[det($\frac{1}{W}\nabla\nabla\chi^2$)], -2log(N!), -Mlog(4$\pi$) and -2log[p($\boldsymbol{\theta}\vert$N)] remain constant as W increases. As such, also the sum of these terms remains constant for more pixels. This sum has been depicted as $\mathcal{O}(1)$ in Fig.~\ref{fig:5}, denoting a term that tends to a constant as W $\rightarrow\infty$. In this way, Eq. (\ref{eq:MAPcriterion2}) can be written for W $\rightarrow\infty$ as
\begin{equation}
\begin{split}
\label{eq:MAPcriterion3}
-2log[p(N\vert\textbf{w})] &= -2log(\hat{L}) + Mlog(W) + \mathcal{O}(1) \\
&\approx -2log(\hat{L}) + Mlog(W)
\end{split}
\end{equation}
implying that the MAP probability rule is asymptotically equivalent with the BIC in Eq. (\ref{eq:BIC}) as -2log($\hat{L}$) and Mlog(W) are the dominant terms when W $\rightarrow\infty$. Interestingly, the behaviour of the Hessian matrix of $\chi^2$($\boldsymbol{\theta}$) evaluated at $\hat{\boldsymbol{\theta}}$, $\nabla\nabla\chi^2$, and of the prior density, p($\boldsymbol{\theta}\vert$N), chosen as a product of uniform distributions for each parameter $\theta_m$, is dependent on the parametrization of the model. Fig.~\ref{fig:5}(a) results from fitting the parameters $\zeta$ and $\eta$ in pixel intensities, depicted in electron counts, and $\rho$, $\beta_x$ and $\beta_y$ in $\AA$. A different parametrization might be to use a normalized model where the parameters can fluctuate between 0 and 1. Fig.~\ref{fig:5}(b) shows the behaviour of the various terms of Eq. (\ref{eq:MAPcriterion2}) of such a model. Here, the terms log[det($\frac{1}{W}\nabla\nabla\chi^2$)] and -2log[p($\boldsymbol{\theta}\vert$N)] have been shifted as compared to Fig.~\ref{fig:5}(a). Yet, they show the same constant behaviour as a function of the number of pixels W. The behaviour of these terms does not necessarily remain constant as a function of W under all parametrizations as shown in Fig.~\ref{fig:5}(c). Here, the parameters $\zeta$ and $\eta$ have been fitted in electron counts and $\rho$, $\beta_x$ and $\beta_y$ in pixels. From Fig.~\ref{fig:5}(c), it is apparent that both log[det($\frac{1}{W}\nabla\nabla\chi^2$)] and -2log[p($\boldsymbol{\theta}\vert$N)] depend on the number of pixels W. Although $\nabla\nabla\chi^2$ and p($\boldsymbol{\theta}\vert$N) are not invariant under reparametrization, yet the term $\mathcal{O}$(1), which is the sum of log[det($\frac{1}{W}\nabla\nabla\chi^2$)], -2log(N!), -Mlog(4$\pi$) and -2log[p($\boldsymbol{\theta}\vert$N)], remains invariant under all parametrizations, as shown in Fig.~\ref{fig:5}. Therefore, the model complexity of Eq. (\ref{eq:complex}) as described by the MAP probability rule remains unchanged under different parametrizations of the model. Moreover, as the goodness of fit is also independent of the model description, the MAP probability rule is invariant under reparametrization.

\subsection{Influence of prior density}\label{prior}
\begin{figure}[ht]
\centering
\includegraphics{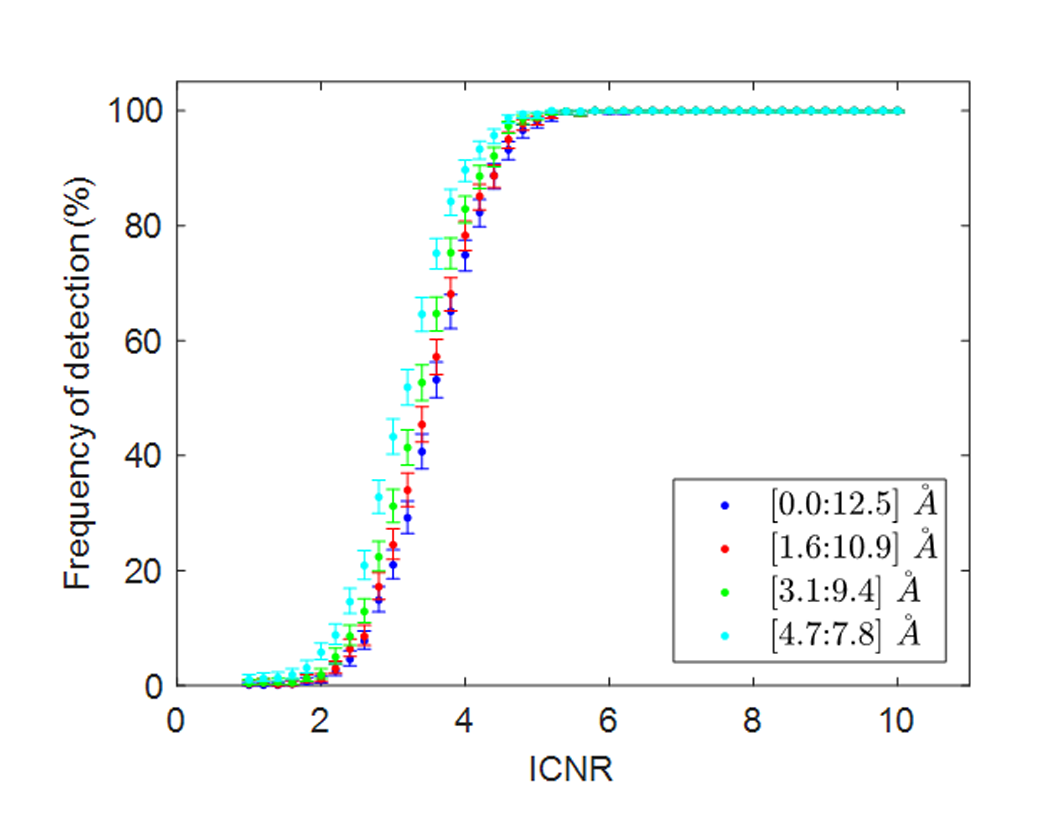}
\caption{Detection rate of a Au atom from simulated HAADF STEM images of 12.5 $\AA$ by 12.5 $\AA$ with a pixel size of 0.25 $\AA$ and an incoming electron dose of 10$^4$ e$^-$/$\AA^2$ as a function of ICNR for different predefined ranges of $\beta_x$ and $\beta_y$.}
\label{fig:6}
\end{figure}

\indent As the prior density p($\boldsymbol{\theta}\vert$N) is part of how the MAP probability rule determines model complexity, as shown in Eq. (\ref{eq:complex}), atom detection in HAADF STEM images depends on the predefined parameter ranges when p($\boldsymbol{\theta}\vert$N) is defined as a product of uniform distributions for each parameter $\theta_m$ individually, as given by Eq. (\ref{eq:prior1}). Ideally, the result of a robust detection method should not depend strongly on the prior. In this section, the influence of different a priori chosen parameter ranges to atom detection is investigated as a function of ICNR since the prior might influence atom detection differently for different ICNR values. For this, HAADF STEM images of 12.5 $\AA$ by 12.5 $\AA$ with a pixel size of 0.25 $\AA$ with different ICNR values have been simulated of an individual Au atom using MULTEM, where the atom is located in the middle of each simulated image, i.e. $\beta_x$ = $\beta_y$ = 6.25 $\AA$. The remaining simulation parameters can be found in Table \ref{tab:parameters}. Each image has been generated 1000 times containing random Poisson noise and the ICNR of the atom in the image has been altered by adding a constant background while keeping the incoming electron dose fixed to 10$^4$ e$^-$/$\AA^2$. For detecting the Au atom from the noise disturbed images by the MAP probability rule, a model assuming the atom to be Gaussian shaped has been used, where a constant background $\zeta$ and width $\rho$, height $\eta$ and x- and y-coordinate $\beta_x$ and $\beta_y$ of the atom need to be estimated. First, the parameters $\zeta$ and $\eta$ have been chosen to range from 0 up to the maximum pixel intensity in the simulated images, whereas the parameters $\rho$, $\beta_x$ and $\beta_y$ range according to the field of view of the image, i.e. from 0 $\AA$ up to 12.5 $\AA$. In order to investigate the effect of a different choice of the predefined parameter ranges, the ranges of $\beta_x$ and $\beta_y$ have been reduced, corresponding to taking into account more and more informative prior knowledge about the location of the Au atom. Fig.~\ref{fig:6} shows that, in general, the detection rate as a function of ICNR is not influenced by different predefined ranges on $\beta_x$ and $\beta_y$ when the ICNR is high. 
For lower ICNR values, though, the detection rate of the Au atom increases for smaller predefined ranges on $\beta_x$ and $\beta_y$. This shows that when more correct prior knowledge can be taken into account, it is beneficial to do so since it increases the chance of detecting atoms from low ICNR STEM images. As compared to the other model-selection criteria considered in this paper, the MAP probability rule offers a more flexible way to detect atoms from HAADF STEM images due to the fact that the prior can be tuned, resulting into a different value for the complexity of the model under consideration. Moreover, by using the MAP probability rule, it is clear what prior knowledge has been taken into account during the analysis, which is not always straightforward for other model-selection criteria.

\section{Conclusions}\label{conclusions}
In the present paper, the methodology of the recently proposed MAP probability rule to detect single atoms from high-resolution HAADF STEM images \cite{fatermans(2018)} has been explained in detail. The method is built upon model-based parameter estimation and by making use of a Bayesian approach it allows for automatic and objective structure quantification. The MAP probability rule is especially useful for the analysis of the structure of beam-sensitive nanomaterials. Typically, images of such materials exhibit low CNR due to the use of a limited incoming electron dose in order to avoid beam damage. Visual inspection of such images is unreliable and might lead to biased results. It has been shown that approximate analytical expressions can be derived for the probability of the presence of a certain number of atomic columns in the image data. The MAP probability rule selects the number of columns that maximizes this probability. \\
\indent Moreover, it has been shown that the MAP probability rule can be effectively used as a tool to evaluate the relation between STEM image quality measures and atom detectability. This has resulted into the introduction of the ICNR as a new image quality measure that better correlates with atom detectability than conventional measures such as SNR and CNR. Atomic columns resulting from images with ICNR values of less than around 5.0 become challenging to accurately detect as typically the detection rate of 100 $\%$ drops rapidly starting from this value. \\
\indent In addition, the relation of the MAP probability rule with model selection has been thoroughly investigated. It has been explained that model-selection criteria are based on a tradeoff between high goodness of fit and low model complexity. Interestingly, the complexity term of the MAP probability rule depends on three dimensions of model complexity, namely the number of parameters, functional form and extension of the parameter space, as opposed to other model-selection criteria, which has been shown to allow for more accurate atom detection from STEM images. Furthermore, the MAP probability rule allows for a clear and flexible incorporation of prior knowledge, which is often not the case for other model-selection methods.

\section*{Acknowledgements}
The authors acknowledge financial support from the Research Foundation Flanders (FWO, Belgium) through project fundings (No. W.O.010.16N, No. G.0368.15N, No. G.0502.18N). This project has received funding from the European Research Council (ERC) under the European Unions Horizon 2020 research and innovation programme (Grant Agreement No. 770887).

\appendix

\section{Analytical derivation of posterior probability}\label{appendixA}
Given the assumption of an a priori equiprobable number of atomic columns N, the evaluation of the MAP probability rule comes down to calculating the marginal likelihood p(\textbf{w}$\vert$N) in Eq. (\ref{eq:marglikelihood}) and determining the number of columns N with the highest posterior probability p(N$\vert\textbf{w}$). If an analytical calculation of the M-dimensional integral in Eq. (\ref{eq:marglikelihood}) is possible, the calculation of the marginal likelihood is significantly facilitated. Approximating the likelihood function p(\textbf{w}$\vert\boldsymbol{\theta}$,N) by a normal distribution, given by Eq. (\ref{eq:normal}), in combination with the choice of a uniformly distributed prior density function p($\boldsymbol{\theta}\vert$N), given by Eq. (\ref{eq:prior1}), allows for an approximate analytical result. Substituting Eq. (\ref{eq:normal}) and Eq. (\ref{eq:prior1}) in Eq. (\ref{eq:marglikelihood}) results into
\begin{equation}
\label{eq:marglikelihoodfull1}
p(N\vert\textbf{w})\propto N!\cdot\Bigg(\prod_{m=1}^M\frac{1}{\theta_{m_{max}}-\theta_{m_{min}}}\Bigg)\int_D\frac{e^{-\chi^2(\boldsymbol{\theta})/2}}{\prod_{k=1}^K\prod_{l=1}^L[2\pi w_{kl}]^{1/2}}d^M\boldsymbol{\theta},
\end{equation}
where
\begin{equation}
\label{eq:domain}
D = \{(\theta_1, ..., \theta_M) \in \mathbb{R}^M \; \text{for} \; m=1, ..., M\text{:} \; \theta_{m_{min}} \leqslant \theta_m \leqslant \theta_{m_{max}}\}.
\end{equation}
The factor N! arises from the number of ways the parameters of the Gaussian peaks can be permuted, as labelling of the N peaks is arbitrary. Therefore, there are N! equivalent maxima of the likelihood function \cite{sivia(2006)}. The expression in Eq. (\ref{eq:marglikelihoodfull1}) can be calculated by expanding the likelihood function by a second order Taylor series around the parameter vector $\hat{\boldsymbol{\theta}}$ that maximizes the likelihood function, which is known as the maximum likelihood estimate:
\begin{equation}
\label{eq:taylor}
\begin{split}
e^{-\chi^2(\boldsymbol{\theta})/2}\approx e^{-\chi^2(\hat{\boldsymbol{\theta}})/2}\times e^{-(\boldsymbol{\theta}-\hat{\boldsymbol{\theta}})^T\bigg[\frac{\partial\chi^2(\boldsymbol{\theta})}{\partial\boldsymbol{\theta}}\big|_{\boldsymbol{\theta}=\hat{\boldsymbol{\theta}}}\bigg]/2}\times e^{-(\boldsymbol{\theta}-\hat{\boldsymbol{\theta}})^T\bigg[\frac{\partial^2\chi^2(\boldsymbol{\theta})}{\partial\boldsymbol{\theta}\partial\boldsymbol{\theta}^T}\big|_{\boldsymbol{\theta}=\hat{\boldsymbol{\theta}}}\bigg](\boldsymbol{\theta}-\hat{\boldsymbol{\theta}})/4}.
\end{split}
\end{equation}
Since $\frac{\partial\chi^2(\boldsymbol{\theta})}{\partial\boldsymbol{\theta}}\big|_{\boldsymbol{\theta}=\hat{\boldsymbol{\theta}}}$ = 0, the second term in Eq. (\ref{eq:taylor}) is equal to one. Then, by writing $\chi^2$($\hat{\boldsymbol{\theta}}$) as $\chi^2_{min}$ and $\frac{\partial^2\chi^2(\boldsymbol{\theta})}{\partial\boldsymbol{\theta}\partial\boldsymbol{\theta}^T}\big|_{\boldsymbol{\theta}=\hat{\boldsymbol{\theta}}}$ as $\nabla\nabla\chi^2$, Eq. (\ref{eq:taylor}) becomes
\begin{equation}
\label{eq:taylorred}
e^{-\chi^2(\boldsymbol{\theta})/2}\approx e^{-\chi^2_{min}/2}\times e^{-(\boldsymbol{\theta}-\hat{\boldsymbol{\theta}})^T\nabla\nabla\chi^2(\boldsymbol{\theta}-\hat{\boldsymbol{\theta}})/4},
\end{equation}
which means that Eq. (\ref{eq:marglikelihoodfull1}) is given by
\begin{equation}
\label{eq:marglikelihoodfull21}
p(N\vert\textbf{w})\propto N! \cdot e^{-\chi^2_{min}/2} \cdot \Bigg(\prod_{m=1}^M\frac{1}{\theta_{m_{max}}-\theta_{m_{min}}}\Bigg)\int_D\frac{e^{-(\boldsymbol{\theta}-\hat{\boldsymbol{\theta}})^T\nabla\nabla\chi^2(\boldsymbol{\theta}-\hat{\boldsymbol{\theta}})/4}}{\prod_{k=1}^K\prod_{l=1}^L[2\pi w_{kl}]^{1/2}}d^M\boldsymbol{\theta}.
\end{equation}
This expression contains a Gaussian multiple integral which can be solved analytically under the assumptions that i) the maximum likelihood estimate $\hat{\boldsymbol{\theta}}$ lies well within the support of the prior density function, described by Eq. (\ref{eq:prior1}), and ii) the likelihood function has only one significant maximum \cite{sivia(1992)}. Then, the integral in Eq. (\ref{eq:marglikelihoodfull21}) is well approximated by an integral over $\mathbb{R}^M$, resulting in the following expression for the posterior probability of the presence of N atomic columns in the image, given the observed image pixel values $\textbf{w}$, for the model described by Eq. (\ref{eq:model}):
\begin{equation}
\label{eq:MAPfinal2}
\begin{split}
p(N\vert\textbf{w})  & \propto \frac{N!}{[(\beta_{x_{max}}-\beta_{x_{min}})(\beta_{y_{max}}-\beta_{y_{min}})(\rho_{max}-\rho_{min})]^N} \\
& \times \frac{e^{-\chi_{min}^2/2}}{(\eta_{max}-\eta_{min})^N(\zeta_{max}-\zeta_{min})}\times\frac{(4\pi)^{M/2}[det(\nabla\nabla\chi^2)]^{-1/2}}{\prod_{k=1}^K\prod_{l=1}^L[2\pi w_{kl}]^{1/2}}.
\end{split}
\end{equation}
This expression is equivalent to Eq. (\ref{eq:MAPfinal}), where the terms which are independent of N have been dropped.

\section{Model complexity of posterior probability}\label{appendixB}
Model selection methods consider a tradeoff between high goodness of fit and low model complexity in order to select the model which most closely describes the underlying process that generated the experimental data. As a model selection criterion, the MAP probability rule is not different. The model complexity can be quantified by writing the analytical expression of p(N$\vert$\textbf{w}) in the form of Eq. (\ref{eq:generalselection}), hereby describing the goodness of fit by p(\textbf{w}$\vert\hat{\boldsymbol{\theta}}$,N) which is given by Eq. (\ref{eq:normal}) and using the prior density p($\boldsymbol{\theta}\vert$N) of Eq. (\ref{eq:prior1}). This results into 
\begin{equation}
\label{eq:complex2}
p(N\vert\textbf{w}) \propto \frac{p(\textbf{w}\vert\hat{\boldsymbol{\theta}},N)}{[det(\nabla\nabla\chi^2)]^{1/2}/[N!(4\pi)^{M/2}p(\boldsymbol{\theta}\vert N)]}
\end{equation}
where the denominator reflects the model complexity as given by Eq. (\ref{eq:complex}). The term $\nabla\nabla\chi^2$ is related to the observed Fisher information matrix $\hat{J}$ \cite{dodge(2003)}, which holds the information that is contained by the observed data about the unknown parameters $\boldsymbol{\theta}$. It is described by the Hessian matrix of minus the logarithm of the likelihood function p(\textbf{w}$\vert\boldsymbol{\theta}$,N), given by Eq. (\ref{eq:normal}), evaluated at the maximum likelihood estimate $\hat{\boldsymbol{\theta}}$. As such,
\begin{equation}
\label{eq:fisher}
\hat{J} = -\frac{\partial^2log[p(\textbf{w}\vert\boldsymbol{\theta},N)]}{\partial\boldsymbol{\theta}\partial\boldsymbol{\theta}^T}\bigg|_{\boldsymbol{\theta}=\hat{\boldsymbol{\theta}}} = \frac{\partial^2[\chi^2(\boldsymbol{\theta})/2]}{\partial\boldsymbol{\theta}\partial\boldsymbol{\theta}^T}\bigg|_{\boldsymbol{\theta}=\hat{\boldsymbol{\theta}}}
\end{equation}
where  $\frac{\partial^2\chi^2(\boldsymbol{\theta})}{\partial\boldsymbol{\theta}\partial\boldsymbol{\theta}^T}\big|_{\boldsymbol{\theta}=\hat{\boldsymbol{\theta}}}$ can be written in short as $\nabla\nabla\chi^2$.

\end{document}